# Performance of Distance Relays in Presence of IPFC

M. Pouyan, F. Razavi, M. Rashidi-Nejad

*Abstract*-- In this paper, the analytical and simulation results of the application of distance relay for the protection of transmission line incorporating Inter-line Power Flow Controller (IPFC) are presented. Firstly a detailed model of the IPFC and its control is proposed and then it is integrated into the 8-bus transmission system for the purposes of accurately simulating the fault transients. The simulation results show the impact of different operational mode of IPFC on the performance of a distance protection relays.

*Index Terms*--Flexible ac transmission system (FACTS), inter-line power flow controller (IPFC), distance protection relay, power system protection.

## I. Nomenclature

| | |
|---|---|
| $P_{se1}$ | Real power injected to ac line by VSC1. |
| $P_{se2}$ | Real power injected to ac line by VSC2. |
| $V_{pq}$ | Injected voltage |
| $V_{1p}$ | Inphase component of injected voltage with respect to ac line current $I_1$ |
| $V_{1q}$ | quadrature component of injected voltage with respect to ac line current $I_1$ |
| $V_{2p}$ | Inphase component of injected voltage with respect to ac line current $I_2$ |
| $V_{2q}$ | quadrature component of injected voltage with respect to ac line current $I_2$ |
| $V_{1m}$ | Injected voltage magnitude of VSC1 |
| $V_{2m}$ | Injected voltage magnitude of VSC2 |

## II. Introduction

THE use of flexible ac transmission system (FACTS) controllers in power system transmission has been of a great interest in recent years for increasing the power transfer capability and enhancing power system controllability and stability[1-4]. However, the implementation of FACTS controllers in transmission systems introduces new power system issues in the field of power system protection. Amongst the different types of FACTS controllers, Inter-line power flow controller (IPFC) is considered to be one of the most effective in the control of power flow. In its general form, the IPFC employs a number of dc to ac converters, each providing series compensation for a different line. The converters are linked together at their dc terminals and connected to the ac systems through their series coupling transformers .With this scheme, in addition to providing series reactive compensation, any converter can be controlled to supply active power to the common dc link from its own transmission line. In the IPFC structure, each converter has the capability to operate a stand-alone SSSC [5-8].

Because of the presence of IPFC controllers in a fault loop, the voltage and current signals at the relay point will be affected in both the steady state and the transient state. This in turn will affect the performance of existing protection schemes, such as the distance relay which is one of the very widely used methods in transmission line protection [9-12].

Some researches have been done to evaluate the performance of a distance relay for transmission systems with FACTS controllers. In [13, 14] the authors have studied the effect of STATCOM on a distance relay at different load levels. The work in [15] has presented a study of the impact of FACTS on the tripping boundaries of distance relay. The work in [16] shows that thyristor-controlled series capacitor (TCSC) has a major influence on the mho characteristic, in particular the reactance and directional characteristic, making the protected region unstable. The study in [17, 18] also shows that the presence of FACTS controllers in a transmission line will affect the trip boundary of a distance relay, and both the parameters of FACTS controllers and their location in the line have an impact on the trip boundary. In [19], the impact of midpoint shunt-FACTS compensated line on the performance of a stand-alone single distance relay has been studied. Reference [20] shows that the series capacitor affects the distance protection and proposes a mitigative method by using new communication-aided schemes. The authors in [21, 22] have studied the effect of UPFC on a distance relay and an apparent impedance calculation procedure based on the power frequency sequence component is then investigated. In [23, 24], the performance of various distance protection schemes for different fault types, fault locations and system conditions, on transmission lines with shunt-FACTS devices applied for midpoint voltage control has been evaluated.

All the studies clearly show that when FACTS controllers are in a fault loop, their voltage and current injections will affect both the steady state and transient components in voltage and current signals, and hence the apparent impedance seen by a conventional distance relay is different from that for a system without FACTS.

Some works, which are mentioned above, have investigated the impact of FACTS devices on distance relays, but none of them have been conducted to investigate the IPFC's impact. The present work is evaluating the performance of distance protection relays in presence of IPFC using PSCAD and is

Mojtaba Pouyan and Farzad Razavi is with Electrical Engineering Department, Tafersh University, Markazi, IRan (e-mail: mojtaba.pouyan@taut.ac.ir) and (email: razavi.farzad@taut.ac.ir)

Masoud Rashidi-Nejad is with Department of Electrical Engineering, Bahonar University, Kerman, Iran. (email: mrashidi@mail.uk.ac.ir)

arranged in following orders. First a model of IPFC system is presented. Then the IPFC control blocks including its control strategies are described which is then embedded into a 150-KV 8-bus transmission system; the simulation results are then employed to study the performance of distance relays in presence of IPFC when a three-phase direct fault ($R_f = 0$) occurs in Zone1 which cover 80% of the relevant line.

### III. IPFC STRUCTURE AND 8_BUS TRANSMISSION SYSTEM

#### A. IPFC Structure

Fig. 1(a) shows the schematic representation of an IPFC. There are two back-to-back voltage-source converters (VSCs), based on the use of gate-turnoff (GTO) thyristor valves. The VSCs produce voltages of variable magnitude and phase angle. These voltages are injected in series with the managed transmission lines via series transformers. The injected voltages are represented by the voltage phasors $V_{1m}$ and $V_{2m}$ in Fig. 1(a). The converters labeled VSC1 (Master) and VSC2 (Slave) in Fig. 1(a) are coupled together through a common dc link. Fig. 1(b) illustrates the IPFC phasor diagram. With respect to the transmission-line current $I_i$ ($i = 1,2$), inphase and quadrature phase components of injected voltage ($V_{1p}, V_{2p}$), respectively, determine the negotiated real and reactive powers of the respective transmission lines. The real power exchanged at the ac terminal is converted by the corresponding VSC into dc power which appears at the dc link as a negative or a positive demand. Consequently, the real power negotiated by each VSC must be equal to the real power negotiated by the other VSC through the dc lines. As shown in Fig. 1(b), VSC1 is operated at point A. Therefore, VSC2 must be operated along the complementary voltage compensation line, such as point B, to satisfy the real power demand of VSC1. This is given by [6, 8]:

$$P_{se1} + P_{se2} = 0$$

#### B. IPFC Control Model

The IPFC control system illustrated in Fig. 2, is based upon PI (Proportional-Integral) controllers. The control system of the IPFC can be divided into two parts: the controls of master and slave VSCs. For the VSC1 (master VSC), the series injected voltages are determined by closed-loop control systems to ensure that the desired active and reactive powers flowing in the master transmission line are maintained. The desired $P_{ref1}$ and $Q_{ref1}$ are compared with the measured active and reactive power flows in the first transmission line, $P_{net1}$ and $Q_{net1}$, and the errors are used to derive the desired direct and quadrature component of the first series inverter voltage, $V_{d1}$ and $V_{q1}$, respectively, through the PI controllers. The magnitude $V_{1m}$ (m I) and phase angle of series converter voltage (Alpha I) can be obtained by a rectangular to polar transformation of $V_{d1}$ and $V_{q1}$ component (Fig. 2 (a)). In order to balance the real power exchange among the series converters and maintain the common dc link voltage, the VSC2 (slave VSC) has one control degree of freedom and can only control the absorption or injection of the reactive power to the second transmission line and control the active power flow of its own line [6]. So desired dc capacitor voltage DC-set-point and $P_{ref2}$ are compared with the measured common dc link voltage and active power flow in the second transmission line, $V_{dc}$ and $P_{net2}$, , and the errors are passes throw a PI controller to produce the desired direct and quadrature component of the second series inverter voltage, $V_{d1}$ and $V_{q1}$, respectively. The magnitude $V_{2m}$ (m II) and phase angle of the second series converter voltage (Alpha II) will be obtained by a rectangular to polar transformation of $V_{d2}$ and $V_{q2}$ component (Fig. 2 (b)).

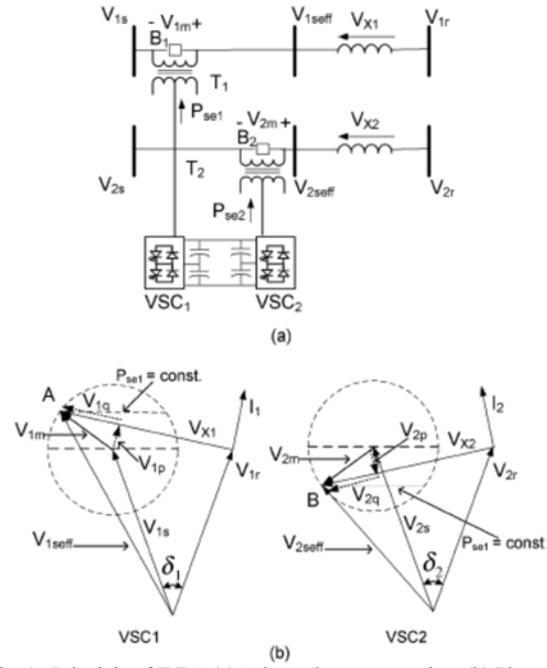

Fig. 1. Principle of IPFC. (a) Schematic representation. (b) Phasor diagram.

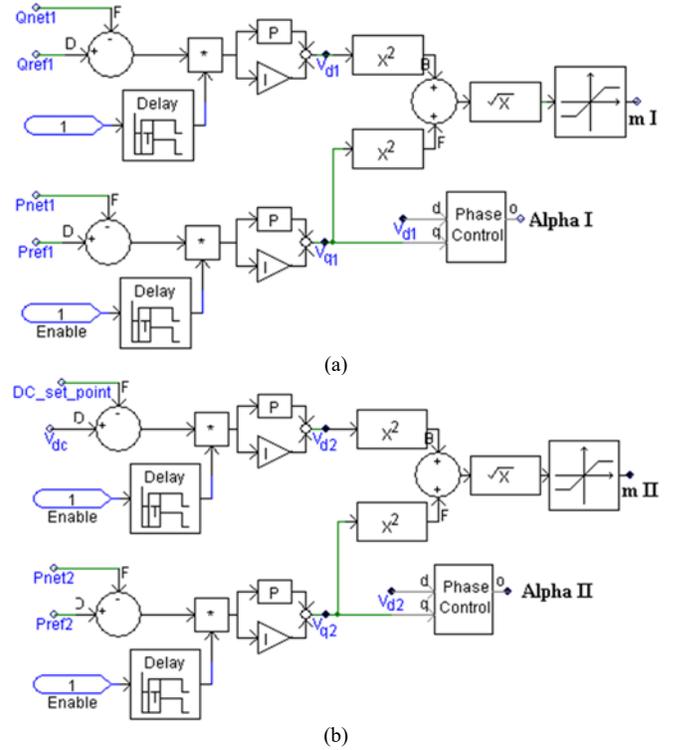

Fig. 2. The controller of VSCs (a): Master VSC (b): Slave VSC

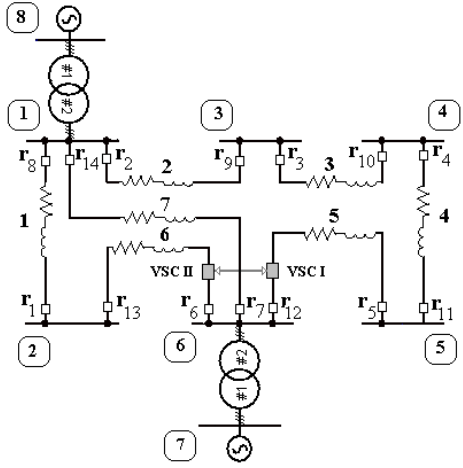

Fig. 3. 8-bus transmission system

TABLE I
LINES'S DATA

| LINE | R (PU) | X (PU) | V (kV) |
|---|---|---|---|
| 1 | 0.0018 | 0.0222 | 150 |
| 2 | 0.0018 | 0.0222 | 150 |
| 3 | 0.0018 | 0.02 | 150 |
| 4 | 0.0022 | 0.02 | 150 |
| 5 | 0.0022 | 0.02 | 150 |
| 6 | 0.0018 | 0.02 | 150 |
| 7 | 0.0022 | 0.0222 | 150 |

TABLE II
GENERATORS' DATA

| Generator | X (p.u.) | V (kV) |
|---|---|---|
|  | 0.1 | 10 |

TABLE III
TRANSFORMERS' DATA

| Transformer | X (p.u.) |
|---|---|
|  | 0.02666 |

### C. Transmission system model

In this study, IPFC is embedded in a standard 8-bus transmission system which is shown in Fig. 3. This network consists of 8 buses, 7 lines, 2 transformers and 2 generators. The data of the network is given in Tables I, II and III. R (p.u.) and X (p.u.) are based on 100 MVA and 150 kV. It is assumed that all the lines are protected by distance relays. All distance relays have impedance characteristic [25].

## IV. APPARENT IMPEDANCE ANALYSIS

For the analysis of the operation of a distance relay in presence on IPFC, the power system shown in Fig. 3 is used; IPFC converters are installed on line 5 and 6. The apparent impedance calculation is based on symmetrical component transformation using power frequency components of voltage and current signals measured at the relay point.

When a fault occurs on the line 5, and the distance $n*L$ from the relay point, the positive, negative and zero sequence networks of the system during the fault are as shown in Fig. 4.

$$V_{1s} = V_{1pq} + nZ_1 I_{1s} + R_f I_{1f} \quad (1)$$
$$V_{2s} = V_{2pq} + nZ_1 I_{2s} + R_f I_{2f} \quad (2)$$
$$V_{0s} = V_{0pq} + nZ_0 I_{0s} + R_f I_{0f} \quad (3)$$

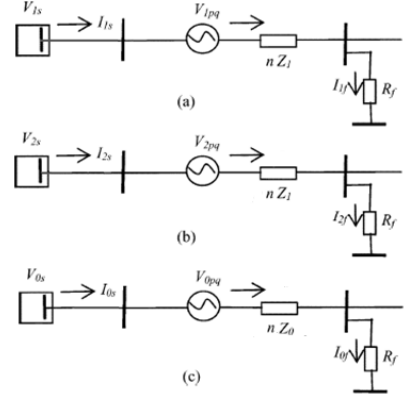

Fig. 4. Sequence networks of the system from the relay location to fault. (a) Positive sequence network. (b) Negative sequence network. (c) Zero sequence network.

where
$V_{1s}, V_{2s}, V_{0s}$ sequence phase voltages at the relay location
$V_{1pq}, V_{2pq}, V_{0pq}$ series sequence phase voltages injected by master VSC of IPFC
$I_{1s}, I_{2s}, I_{0s}$ sequence phase currents at the relay location
$I_{1f}, I_{2f}, I_{0f}$ sequence phase currents in the fault
$Z_1, Z_0$ sequence impedance of the transmission line
$n$ per-unit distance of a fault from the relay location.

From above, the voltage at the relay point can be derived as:
$$V_s = nZ_1 I_s + n(Z_0 - Z_1)I_{0s} + V_{pq} + R_f I_f \quad (4)$$

where
$$V_s = V_{1s} + V_{2s} + V_{0s} \quad (5)$$
$$V_{pq} = V_{1pq} + V_{2pq} + V_{0pq} \quad (6)$$
$$I_s = I_{1s} + I_{2s} + I_{0s} \quad (7)$$
$$I_f = I_{1f} + I_{2f} + I_{0f} \quad (8)$$

In the transmission system without IPFC, for a single phase-to-ground fault, the apparent impedance of distance relay can be calculated using the equation

$$Z = V_R / \left(I_R + \frac{Z_0 - Z_1}{Z_1} I_{0R}\right) = \frac{V_R}{I_{relay}} \quad (9)$$

where
$V_R, I_R$ phase voltage and current at relay point
$I_{0R}$ zero sequence phase current
$I_{relay}$ relaying current

If this traditional distance relay is applied to the transmission system with IPFC, the apparent impedance seen by this relay can be expressed as:

$$Z = \frac{V_s}{I_s + \frac{Z_0 - Z_1}{Z_1} I_{0s}} = \frac{V_s}{I_{relay}} = nZ_1 + \frac{V_{pq}}{I_{relay}} + \frac{R_f}{I_{relay}} I_f \quad (10)$$

In our study $R_f$ is assumed to be zero. So, the apparent impedance seen by this relay can be presented by:

$$Z = nZ_1 + \frac{V_{pq}}{I_{relay}} \quad (11)$$

The injected voltage is equal to injected impedance cross the relay current:

$$V_{pq} = Z_{pq}I_{relay} \quad (12)$$

where

$Z_{pq}$      injected impedance

therefore the apparent impedance seen by relay can be attained by:

$$Z = nZ_1 + Z_{pq} = nZ_1 + (R_{pq} + jX_{pq}) \quad (13)$$

where

$R_{pq}$      injected resistance
$X_{pq}$      injected reactance

It can be seen when the conventional distance relay is applied to the transmission system employing IPFC during the fault, the apparent impedance seen by this relay has two parts: positive sequence impedance from the relay point to fault point, which is what the distance relay is set to measure; the second is due to the impact of IPFC on the apparent impedance.

The foregoing analysis illustrates the effect of IPFC on apparent impedance and hence on the performance of the distance relay.

## V. THE IMPACT OF IPFC ON DISTANCE RELAY

As mentioned before, the VSCs can exchange both the active and reactive power respectively to the transmission line and consequently, control the active and reactive powers flow in the transmission lines. There are two operation modes; the first is when the IPFC injects/absorbs reactive power; the second is when it injects/absorbs active power to/from transmission line. Some loads that not shown here are assumed and the apparent impedance trajectory seen by the distance relay $r_{12}$ are obtained, with and without IPFC, when 3-phase direct fault occurs at 80% of line 5. In Fig. 5 the impedance trajectory chart is depicted as the fault occurs in absence of IPFC.

### A. Reactive power injection/absorption

In this mode, IPFC operates like series capacitor/inductance which in turn decrease/increase the apparent reactance seen by distance relay and cause intangible changes in the apparent resistance (Fig. 6-7) and consequently, increase/decrease the active power flow in the first transmission line. Therefore, in the case of reactive power injection there is a possibility of the distance relay over-reaching which is clearly undesirable (Fig. 8) and vice versa, in the case of reactive power absorption, the apparent Impedance increases which in turn leads to under-reaching of distance relay which is shown in Fig. 9 (same fault condition and location).

### B. Active power injection/absorption

When the master or slave VSC of IPFC absorb/inject active power, they acts as a positive/negative resistance and hence increase/decrease the apparent resistance seen by the distance relay which is depicted in Fig. 10 and cause slite changes in the apparent reactance (Fig. 11). So, in the case of active power injection, the apparent impedance seen by distance

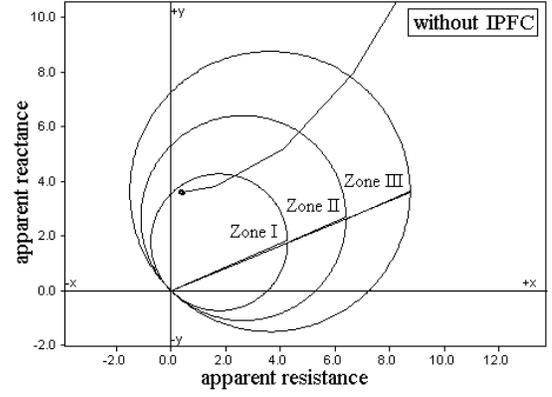
Fig. 5. Apparent impedance without IPFC

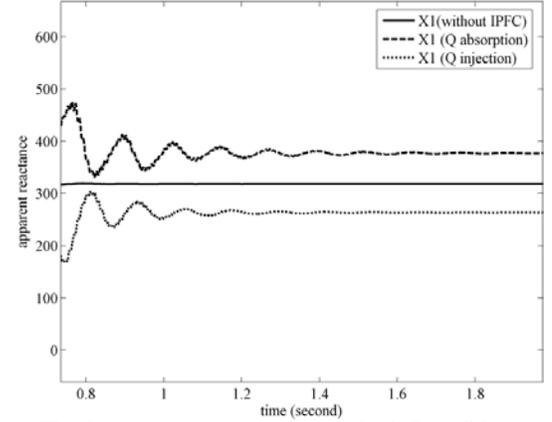
Fig. 6. Apparent reactance seeb by relay in 3 conditions

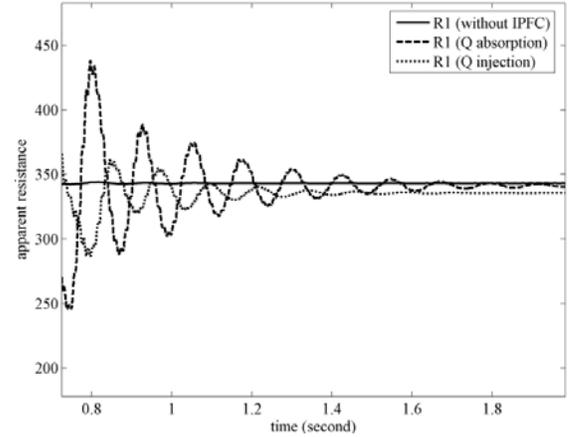
Fig. 7. Apparent resistance seen by relay in 3 conditions

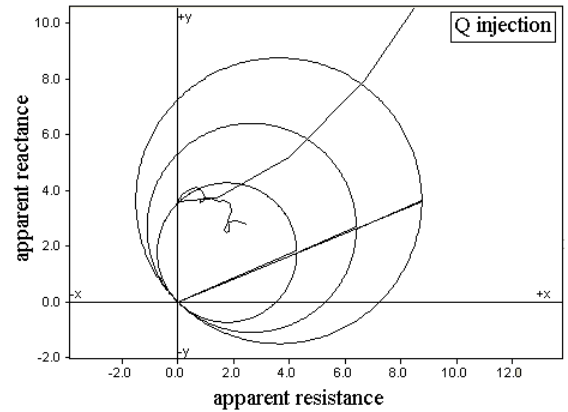

Fig. 8. Apparent Impedance trajectory seen by relay in the case of Q injection

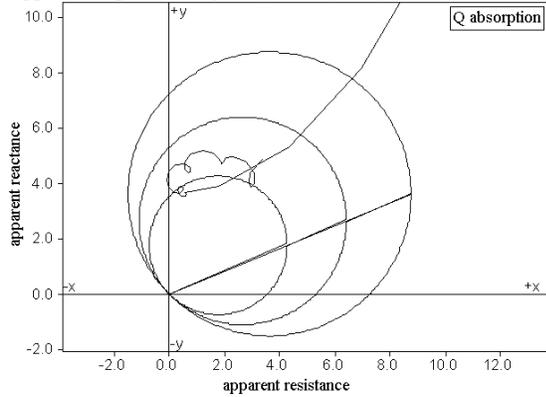

Fig. 9. Apparent Impedance seen by relay in the case of Q absorption

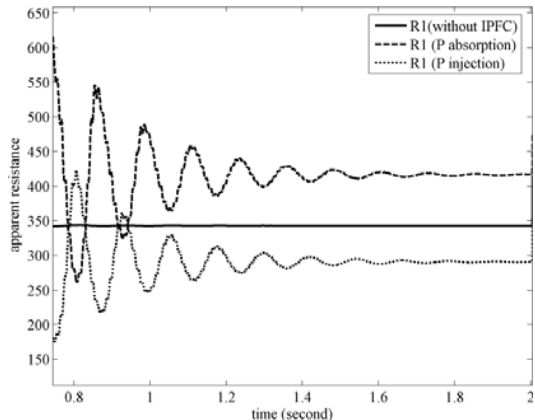

Fig. 10. Apparent resistance seen by relay in 3 conditions

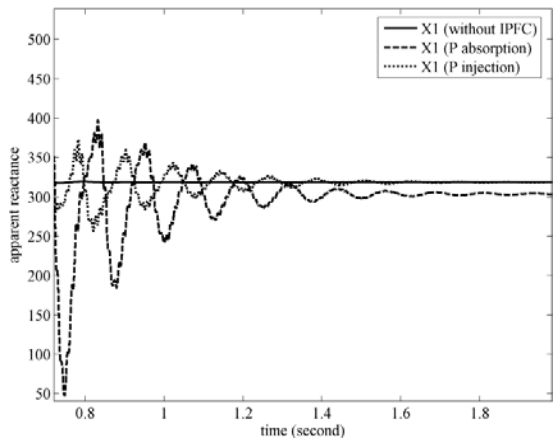

Fig. 11. Apparent reactance seen by relay in 3 conditions

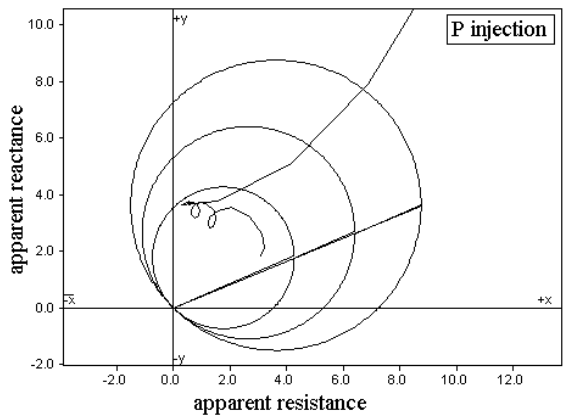

Fig. 12. Apparent Impedance seen by relay in the case of P injection

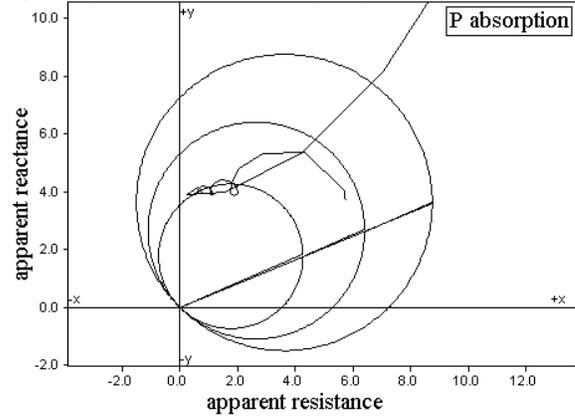

Fig. 13. Apparent Impedance seen by relay in the case of P absorption

realy is reduced and there is a tendency for the relay to over-reach and in the case of active power consumption, the apparent Impedance seen by relay is increased that will lead to the underreaching of distance relay as evident from Figs. 12 and 13, respectively.

## VI. CONCLUSION

In this paper, performance of distance relays in presence of IPFC as integrated into an 8-bus system is presented and analyzed. The results can be inferred as fallow:

When an IPFC injects reactive power into the transmission line, it operates like a capacitor and reduces the apparent impedance seen by distance relay and can cause the distance relay to over-reach.

If it consumes reactive power, it operates like a series inductance which in turn gives rise to the apparent impedance seen by distance relay and hence a possibility of the relay under-reaching.

In the case active power injection, it operates like a negative resistance which decreases the apparent resistance and the apparent impedance seen by the distance relay which will result in the over-reaching of distance relay.

Another issue is the active power absorption which leads to apparent resistance and impedance increment that cause the distance relay to under-reach.

Presenting a new relay setting to overcome the over-reaching and under-reaching phenomena, is a part of on-going work.

## VII. REFERENCES

[1] M.Tahan, T. Hu, "High Performance Multiple String LED Driver with Flexible and Wide Range PWM Dimming Capability," IEEE Applied Power Electronics Conference and Exposition (APEC), Tampa, FL, 26-30 March, 2017.

[2] N. G. Hingorani and L. Gyugyi, Understanding FACTS: Concepts &Technology of Flexible AC Transmission Systems. New York: Wiley, 1999.

[3] M. Tahan and T. Hu, "Multiple String LED Driver With Flexible and High-Performance PWM Dimming Control," IEEE Transactions on Power Electronics, vol. 32, no. 12, pp. 9293-9306, 2017.

[4] M. Mathur and R. K. Varma, Thyristor-Based FACTS Controllers for Electrical Transmission Systems. New York: Wiley, 2002.

[5] Bamgboje, D.O.; Harmon, W.; Tahan, M.; Hu, T. "Low Cost High Performance LED Driver Based on a Self-Oscillating Boost Converter" IEEE Transactions on Power Electronics, vol. 34, no. 10, pp. 10021–10034, 2019.

[6] Gyugyi L, Sen K. K., Schauder C. D. "The Interline Power Flow Controller Concept: A New Approach to Power Flow Management in Transmission Systems" IEEE. Vol 14 No 3, pp 1115-1123, July 1999.

[7] B. Vahidi, S. Jazebi, H.R. Baghaee, M.Tahan,M. R. Asadi, M. R., "Power System Stabilization Improvement by Using PSO-based UPFC", Joint International

Conference on Power System Technology and IEEE Power India Conference, New Delhi, 2008, pp. 1-7.

[8] Moghadasi S.M., Kazemi A., Fotuhi M., Edris A., "Composite System Reliability Assessment Incorporating an Interline Power-Flow Controller" IEEE Transactions on Power Delivery, VOL. 23, NO. 2, pp 1191-1199, Apr. 2008.

[9] H. Wang, M. Tahan, and T. Hu, "Effects of rest time on equivalent circuit model for a li-ion battery," 2018 Annual American Control Conference (ACC), pp. 3101–3106, Boston, MA, 2016.

[10] A. G. Phadke, T. Hlibka, and M. Ibrahim, "Fundamental basis for distance relaying with symmetrical components," IEEE Trans. Power App. Syst., vol. PAS-96, pp. 635–646, Mar./Apr. 1977.

[11] M. Tahan and T. Hu, "Speed-sensorless vector control of surface mounted PMS motor based on modified interacting multiple-model EKF," 2015 IEEE International Electric Machines & Drives Conference (IEMDC), Coeur d'Alene, ID, 2015, pp. 510-515.

[12] D. L. Waikar, S. Elangovan, and A. C. Liew, "Fault impedance estimation algorithm for digital distance relaying)," IEEE Trans. Power Delivery, vol. 9, no. 3, pp. 1375–1383, Jul. 1994.

[13] M. Tahan, D. Bamgboje and T. Hu, "Hybrid control system in an efficient LED driver",2018 Annual American Control Conference (ACC), Milwaukee, WI, 2018.

[14] K. El-Arroudi, G. Joos, and D. T. McGillis, "Operation of impedance protection relays with the STATCOM," IEEE Trans. Power Del., vol. 17, no. 2, pp. 381–387, Apr. 2002.

[15] P. K. Dash, A. K. Pradhan, G. Panda, and A. C. Liew, "Adaptive relay setting for flexible AC transmission systems (FACTS)," IEEE Trans. Power Del., vol. 15, no. 1, pp. 38–43, Jan. 2000.

[16] W. G.Wang, X. G. Yin, J. Yu, X. Z. Duan, and D. S. Chen, "The impact of TCSC on distance protection relay," in Proc. Int. Conf. Power System Technology (POWERCON '98), vol. 1, pp. 18–21, Aug. 1998.

[17] M. Tahan, H. Monsef and S. Farhangi, "A new converter fault discrimination method for a 12-pulse high-voltage direct current system based on wavelet transform and hidden markov models," Simulation, vol. 88, no. 6, pp. 668-679, 2012.

[18] M. Khederzadeh, "The impact of FACTS device on digital multifunctional protective relays," in Proc. IEEE/PES Transmission and Distribution Conf. and Exhib. 2002: Asia Pacific, vol. 3, Oct. 6–10, 2002, pp. 2043–2048.

[19] T. S. Sidhu, R. K. Varma, P. K. Gangadharan, F. A. Albasri, and G. R. Ortiz, "Performance of distance relays on shunt—FACTS compensated transmission lines," IEEE Trans. Power Del., vol. 20, no. 3, pp. 1837–1845, Jul. 2005.

[20] T. S. Sidhu and M. Khederzadeh, "Series Compensated Line Protection Enhancement by Modified Pilot Relaying Schemes" IEEE Trans. Power Del., vol. 21, no. 3, pp. 1191-1198, Jul. 2006.

[21] M.Tahan, H.Monsef, "HVDC Converter Fault Discrimination using Probabilistic RBF Neural Network Based on Wavelet Transform", 4th power systems protections and control conference (PSPC), Tehran, Iran, 2010.

[22] X. Zhou, H. Wang, R. K. Aggarwal and P. Beaumont, "Performance Evaluation of a Distance Relay as Applied to a Transmission System With UPFC" IEEE Trans. Power Del., vol. 21, no. 3, pp. 1137-1147, Jul. 2006.

[23] F. A. Albasri, T. S. Sidhu, and R. K. Varma, "Performance Comparison of Distance Protection Schemes for Shunt-FACTS Compensated Transmission Lines" IEEE Trans. Power Del., vol. 22, no. 4 pp. 2116-2125, Oct. 2007.

[24] M. Tahan, D. Bamgboje and T. Hu, "Flyback-Based Multiple Output dc-dc Converter with Independent Voltage Regulation," 2018 9th IEEE International Symposium on Power Electronics for Distributed Generation Systems (PEDG), Charlotte, NC,2018, pp. 1-8.

[25] Farzad Razavi, Hossein Askarian Abyaneha, Majid Al-Dabbagh, Reza Mohammadia, Hossein Torkaman, "A new comprehensive genetic algorithm method for optimal overcurrent relays coordination" Electric Power Systems Research, pp 713-720,2008.

**Mojtaba Pouyan** was born in 1983. He received B.S degree in electrical engineering from Shahid Bahonar University, Kerman, Iran. Now he is working toward his M.S. degree in electric engineering at Tafresh University. His area of interest is distribution system, optimization techniques, FACTS devices and protection systems.

**Farzad Razavi** received his PhD degree in electrical engineering from Amir-Kabir university, Tehran, Iran. He is currently an assistant professor at Electrical Eng. Dept. Tafresh University, Tafresh, Iran. His area of interest is Flexible Ac Transmission Systems (FACTS) and protection systems.

**Masoud Rashidinejad** received his B.S. degree in Electrical Engineering and M.S. degree in Systems Engineering Isfahan University of Technology, Iran. He received his PhD in Electrical Eng. from Brunel University, London, UK, 2000. He is currently associated professor at Electrical Eng. Dept., Shahid Bahonar University of Kerman, Kerman, Iran. His area of interests is power system optimization, Distribution networks, electricity restructuring and energy management.